\documentclass[sigconf]{acmart}
\usepackage{graphicx}
\AtBeginDocument{%
  \providecommand\BibTeX{{%
    \normalfont B\kern-0.5em{\scshape i\kern-0.25em b}\kern-0.8em\TeX}}}

\setcopyright{acmcopyright}
\copyrightyear{2024}
\acmYear{2024}
\acmDOI{XXXXXXX.XXXXXXX}

\acmConference[SIGGRAPH 2024 Poster ]{SIGGRAPH 2024 Poster}{Aug, 2024}{}
\acmPrice{15.00}
\acmISBN{978-1-4503-XXXX-X/18/06}

\begin{document}

\title{AI in Your Toolbox: A Plugin for Generating Renderings from 3D Models
}

\author{Mingming Wang}
\affiliation{%
  \institution{USC $\&$ TEI Co.}
  \country{}
  }
\email{wangmm913@gmail.com}

\renewcommand{\shortauthors}{Mingming Wang}

\begin{abstract}

With the rapid development of LLMs and AIGC technology, we present a Rhino platform plugin utilizing stable diffusion technology. This plugin enables real-time application deployment from 3D modeling software, integrating stable diffusion models with Rhino's features. It offers intelligent design functions, real-time feedback, and cross-platform linkage, enhancing design efficiency and quality. Our ongoing efforts focus on optimizing the plugin to further advance AI applications in CAD, empowering designers with smarter and more efficient design tools. Our goal is to provide designers with enhanced capabilities for creating exceptional designs in an increasingly AI-driven CAD environment.

\end{abstract}



\begin{teaserfigure}
  \includegraphics[width=\textwidth]{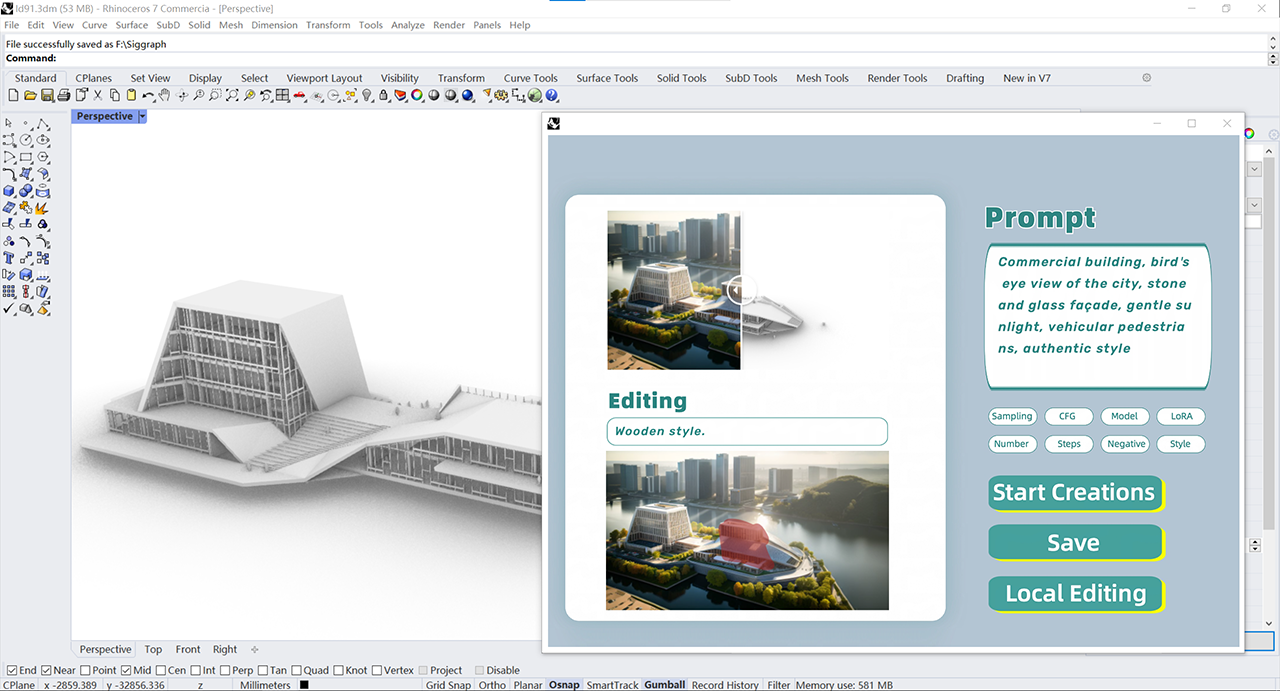}
  \caption{We developed a Rhino platform plugin based on pre-trained stable diffusion model, supporting the generation and editing of architectural renderings from 3D software. The plugin enables cross-platform linkage, providing architects with a more efficient workflow. Leveraging stable diffusion models, we can rapidly generate diverse architectural models and support detailed editing and optimization by user-provided prompts, making the design process more flexible and intuitive. Whether it's preliminary sketches or final architectural renderings, our plugin meets the needs of designers.}
  \Description{}
  \label{fig:teaser}
\end{teaserfigure}

\maketitle

\section{Introduction}

With the rapid development of large language models and image technology~\cite{rombach2022high,saharia2022photorealistic}, the field of Computer-Aided Design (CAD)~\cite{kennard1969computer,kalay2004architecture,barnhill2014computer,barnhill1985surfaces,roudsari2013ladybug,groover1983cad} has also experienced rapid growth. Over the past few decades, CAD software has made significant advancements in functionality and performance, becoming an indispensable tool in modern design and engineering fields. However, with the continuous evolution of artificial intelligence technology, the CAD domain is facing new challenges and opportunities.

The application of artificial intelligence technology has brought many innovations to the CAD field, including new design methods and tools based on machine learning and deep learning. These technologies not only accelerate the design process but also improve the accuracy and efficiency of design. Especially in the field of architectural design, the application of artificial intelligence is changing the way designers work, providing them with more inspiration and creativity. Against this backdrop, we have developed a Rhino platform plugin based on stable diffusion, aiming to utilize artificial intelligence technology to achieve real-time deployment of applications from 3D modeling software. This plugin combines stable diffusion algorithms with the powerful features of the Rhino platform, providing designers with a new design tool that allows them to more flexibly engage in architectural design and engineering planning.

As shown in Fig. 1, our plugin has several prominent features, including intelligent design functions, real-time deployment of applications, and cross-platform linkage. Firstly, we employ stable diffusion algorithms, which can rapidly generate diverse architectural models and support detailed editing and optimization. Secondly, our plugin enables real-time deployment of applications from 3D modeling software, allowing designers to receive immediate feedback and make adjustments during the modeling process, significantly improving design efficiency and quality. Finally, our plugin achieves cross-platform linkage, providing comprehensive support for designers' work. Whether it's preliminary sketches or final architectural renderings, our plugin meets the needs of designers, bringing them a whole new design experience.

In the future, we will continue to focus on optimizing and improving the plugin, continuously promoting the application of artificial intelligence technology in the CAD field, providing designers with more intelligent and convenient design tools, and helping them create more excellent design works.

\section{Applications}

\subsection{Rhino Platform}

Rhinoceros or Rhino, is a powerful and versatile 3D modeling software widely used across various industries, including architecture, industrial design, engineering, and multimedia. Developed by Robert McNeel $\&$ Associates, Rhino offers a robust set of tools for creating, editing, analyzing, and rendering 3D models with unprecedented flexibility and precision.

One of Rhino's key strengths lies in its intuitive and user-friendly interface, which allows designers and architects to quickly bring their ideas to life in a virtual 3D environment. Its extensive suite of modeling tools enables the creation of complex shapes, surfaces, and geometries, ranging from simple primitives to intricate organic forms.

Rhino's open architecture and extensive plugin ecosystem further extend its capabilities, allowing users to customize and enhance the software to suit their specific needs. This flexibility makes Rhino a preferred choice for professionals seeking tailored solutions for their design workflows. 
Considering these features, we have chosen to develop an interface for interoperability with AIGC algorithms on the Rhino platform.

\subsection{Technical Settings}

In our developed plugin, one of the core functionalities is Text-to-Image (T2I) generation. This technology has been extensively showcased in our previous work, particularly in the field of architectural design~\cite{li2023sketch,li2024generating}. By flexibly capturing different perspective images from 3D modeling software and combining them with a series of parameter settings, users can generate a large number of images with varying styles by simply clicking "Start Generation". Our plugin integrates the ControlNet~\cite{zhang2023adding,hu2021lora} model, allowing users to control the generation process by leveraging detailed edge and depth maps as constraints. Additionally, we have incorporated training from the LoRA~\cite{hu2021lora} model, enabling users to ensure stylistic consistency in the generated results based on custom image styles. The combination of these two technologies further enhances the quality and consistency of the generated images, enabling users to more easily control the style and features of the images. In addition to these functionalities, our plugin also offers adjustable diffusion model parameters, allowing users to fine-tune according to their specific needs. Furthermore, image editing techniques~
\cite{ruiz2023dreambooth,li2023layerdiffusion,zhang2023adding,brooks2023instructpix2pix,mokady2023null,balaji2022ediff} based on large models have developed rapidly, we introduced a local editing algorithm, enabling users to regenerate unsatisfactory local areas based on text descriptions. This functionality allows users to achieve fine-tuning and modifications tailored to their specific requirements, ultimately obtaining satisfactory design results. The uniqueness of this plugin lies in its flexibility and controllability. Users can not only generate images through simple operations but also deeply customize according to their creativity and requirements. Whether adjusting the overall style or fine-tuning local details, our plugin provides comprehensive support to help users realize their design visions.

\section{Conclusion}

Our development of a Rhino platform plugin based on stable diffusion models represents a significant step forward in leveraging artificial intelligence to enhance the design process. By combining stable diffusion algorithms with the powerful features of the Rhino platform, our plugin provides designers with intelligent design functions, real-time application deployment, and cross-platform linkage, revolutionizing their approach to architectural design and engineering planning.
Moving forward, our focus remains on refining and enhancing the plugin to further promote the application of artificial intelligence in the CAD field. We are committed to providing designers with more intelligent and convenient design tools, empowering them to create exceptional design works with efficiency and creativity. Through continuous innovation and improvement, we aim to shape the future of CAD design, driving forward the convergence of artificial intelligence and design excellence.

\bibliographystyle{ACM-Reference-Format}
\bibliography{sample-base}

\end{document}